\documentclass{ws-mpla}   %4-4-07

\usepackage{url}

\usepackage{hyperref}
\usepackage{overcite}

\usepackage{graphicx}
\usepackage{amsmath}
\usepackage{slashed}
\usepackage[usenames]{color}
\usepackage{longtable}

\def\beq{\begin{equation}}
\def\eeq{\end{equation}}
\def\bea{\begin{eqnarray}}
\def\eea{\end{eqnarray}}
\def\D0{D\O }

\begin{document}
\newcommand{\mx} {\ensuremath{m_{X}}\xspace}
\newcommand{\mxqsq} {\ensuremath{(m_{X}, q^2)}\xspace}
\newcommand {\pplus}  {\ensuremath{P_{+}}\xspace}
\newcommand{\elsmax}{\ensuremath{(E_{\ell},s_{\mathrm{h}}^{\mathrm{max}})}}
\newcommand{\smax}{\ensuremath{s_{\mathrm{h}}^{\mathrm{max}}}}
\newcommand{\el} {\ensuremath{E_{\ell}}\xspace}
\newcommand {\mb}{\ensuremath{m_b}}
\newcommand {\mc}{\ensuremath{m_c}\xspace}
\newcommand{\btou}{\ensuremath{\bar{B} \to X_u\, l\,  \bar{\nu}_l}}
\newcommand{\btoc}{\ensuremath{\bar{B}\to X_c\, l\,  \bar{\nu}_l}}
\newcommand{\bsg}{\ensuremath{\bar{B}\to X_s\, \gamma}}
\newcommand{\Vcb} {\ensuremath{|V_{cb}|}}
\newcommand{\Vub} {\ensuremath{|V_{ub}| }}
\newcommand{\Vtb} {\ensuremath{|V_{tb}| }}

\def\st{\scriptstyle}
\def\sst{\scriptscriptstyle}
\def\mco{\multicolumn}
\def\epp{\epsilon^{\prime}}
\def\vep{\varepsilon}
\def\ra{\rightarrow}
\def\ppg{\pi^+\pi^-\gamma}
\def\vp{{\bf p}}
\def\ko{K^0}
\def\kb{\bar{K^0}}
\def\al{\alpha}
\def\ab{\bar{\alpha}}
\def\be{\begin{equation}}
\def\ee{\end{equation}}
\def\beq{\begin{equation}}
\def\eeq{\end{equation}}
\def\bea{\begin{eqnarray}}
\def\eea{\end{eqnarray}}
\def\CPbar{\hbox{{\rm CP}\hskip-1.80em{/}}}
\newcommand{\lsim}{
\mathrel{\hbox{\rlap{\hbox{\lower4pt\hbox{$\sim$}}}\hbox{$<$}}}}

\thispagestyle{plain}

\def\bib{B\kern-.05em{I}\kern-.025em{B}\kern-.08em}
\def\btex{B\kern-.05em{I}\kern-.025em{B}\kern-.08em\TeX}

\title{Determination of the CKM matrix elements $|V_{xb}|$}

\markboth{}{}

\author{\footnotesize GIULIA RICCIARDI}

\address{Dipartimento di Fisica, Universit\`a  di Napoli Federico II \\
Complesso Universitario di Monte Sant'Angelo, Via Cintia,
I-80126 Napoli, Italy\\
and \\
 INFN, Sezione di Napoli\\
Complesso Universitario di Monte Sant'Angelo, Via Cintia,
I-80126 Napoli, Italy\\
giulia.ricciardi@na.infn.it}

\maketitle

\begin{abstract}
We  review the current status of  $|V_{cb}|$,  $|V_{ub}|$ and $|V_{tb}|$, the absolute values of the  matrix elements in the CKM  third column.

%\keywords{QCD; heavy flavour; B decays.}
\end{abstract}
\section{Introduction}

In the Standard Model (SM), the strength of weak charged  current interactions of quarks  is codified inside the Cabibbo-Kobayashi-Maskawa (CKM) matrix, but
the actual  values of the CKM matrix   elements  are not
predictable within the SM.
It is  important to ascertain such values precisely, since the  departure of the CKM matrix   from the unit matrix is at the origin of flavour and CP violating processes in the SM.
We briefly  review recent progress in the determination of  $|V_{cb}|$,  $|V_{ub}|$ and $|V_{tb}|$, that is the absolute values of the  matrix elements in the CKM  third column   (see Fig. (\ref{CKM-fig})).
\begin{figure*}[ph]
\bea
%V_{\rm{CKM}}=
\left(%
\begin{array}{ccc}
  V_{ud} & V_{us} &{{\color{red}  V_{ub}}} \\
  V_{cd} & V_{cs} & {{ \color{red} V_{cb}} }\\
  V_{td} & V_{ts} & {\color{red} V_{tb}} \\
\end{array}%
\right)
\nonumber
\eea
\label{CKM-fig}
\caption{The CKM matrix}
\end{figure*}

These matrix elements only contribute to  weak transitions involving the heavy $b$-quark and it is a theoretical advantage the possibility to exploit, in most cases,  the setting of the heavy flavour effective theory.
Another advantage  is that they can be  determined directly  by analyzing  tree-level decays only.
The exchange of a new physics (NP) particle is strongly constrained at tree level. A clean  determination of CKM parameters from tree level processes is  therefore a valuable input for other NP more sensitive estimates. Conversely, relations between $|V_{cb}|$,  $|V_{ub}|$ and other parameters, such as, e.g. the  $\epsilon_K$ dependency on $|V_{cb}|^4$,  can  be exploited to estimate their   values, within or beyond the SM \cite{Buras:2013raa}.

$|V_{cb}|$ and $|V_{ub}|$   play a considerable role in the
analysis of the unitarity triangle. The so-called unitarity
clock, the circle around the origin in the $\bar \rho-\bar \eta$
plane\cite{Buras:2001pn}, is proportional to the ratio $|V_{ub}/V_{cb}|$, and  $|V_{cb}|$ normalizes the whole unitarity triangle.
Most precise determinations of  $|V_{cb}|$ and $|V_{ub}|$ are currently inferred from semi-leptonic decays, exclusive and inclusive ones. In
 Sect.\ref{kin22} we outline the theoretical approaches and set the notations for semi-leptonic decays.
 Sections \ref{vcb}, \ref{vub}, \ref{vtb} report in detail on the determinations of  $|V_{cb}|$, $|V_{ub}|$ and $|V_{tb}|$,  respectively.

%In the SM with 3 quark generations, the top quark is expected to decay to a $W$ boson and a $b$ quark roughly 99.8 \% of the time.

\section{Semi-leptonic decays -  Outline}
\label{kin22}

At the moment, the most precise values of  $|V_{cb}|$ and $|V_{ub}|$ are  inferred from semi-leptonic decays, exclusive and inclusive ones.
Data is provided by electron-positron machines, as LEP and CLEO, but above all by the dedicated Beauty ($B$-) Factories, BaBar and Belle, which have greatly reduced the  errors on previous branching ratio determinations, allowing to attain an unprecedented  high level of precision.
In the $B$-factories, a $\bar B$-$B$  pair is produced
nearly at rest in the $\Upsilon (4S)$  frame and the $\bar B$-$B$   production accounts for approximately 1/4 of
the $e^+ e^- \to$ hadrons cross-section.
The decay products of the $B$ mesons overlap, and the neutrino
from the semileptonic $B$ decay goes undetected. As a result, in order to unambiguously associate  hadrons
with a semileptonic $B$ decay,  the second $B$ meson in the event need to be fully reconstructed.

Semileptonic exclusive $B$ decays are also studied at hadron colliders. However, the
measurements of $|V_{cb}|$ and $|V_{ub}|$ imply the reconstruction, in the $b$-hadron rest frame, of observables difficult to measure at hadron colliders,  such as the
squared invariant mass of the lepton pair $q^2$. At LHCb,  it is possible to improve the $q^2$ resolution
 by exploiting the separation between primary and secondary
vertices, determining the $B$ flight direction vector and measuring
the neutrino momentum with a two-fold ambiguity \cite{Bozzi:2013aba}.
With about 1.2 million $ B^0 \to D^{\ast +} \mu \, \nu$ decays reconstructed in 1 fb$^{-1}$, it  is worthwhile to explore the
LHCb potential  for the $|V_{cb}|$ determination.

 The inclusive and exclusive determinations of $|V_{cb}|$ and $|V_{ub}|$  rely on
different theoretical calculations, each with different (independent) uncertainties, and on
different experimental techniques which have, to a large extent, uncorrelated
statistical and systematic uncertainties. This independence makes
the comparison of $|V_{cb}|$ and $|V_{ub}|$ determination from inclusive and exclusive decays a
powerful test of our physical understanding. In Sects. \ref{subsectionExclusive decays} and \ref{subsectionInclusive decays} we summarize the main points of the theoretical approach and set notations. \footnote{For recent, dedicated reviews to semi-leptonic decays see also Refs. \refcite{Ricciardi:2012pf}, \refcite{Ricciardi:2012dj}, \refcite{Ricciardi:2013jf}.}

\subsection{Exclusive decays}
\label{subsectionExclusive decays}

Let us consider a generic semi-leptonic decay $H \to P l \nu$, where $H$ and $P$
denote a heavy and a light pseudoscalar meson, respectively.
The transition $H \to P$ is mediated by the vector current $V^\mu$ and the hadronic matrix element between the initial and final state can be decomposed in a Lorentz covariant form  built from the independent four-momenta of the decay
\beq
\langle P(p_P)| V^\mu | H(p_H) \rangle = f_+(q^2) \left (p_H^\mu+p_P^\mu-\frac{m^2_H-m_P^2}{q^2} q^\mu \right) +
 f_0(q^2) \frac{m^2_H-m_P^2}{q^2} q^\mu
 \label{LC1}
\eeq
The two
form factors $f_+(q^2)$ and $f_0(q^2)$ depend only on $ q^\mu \equiv p_H^\mu - p_P^\mu$,   the momentum transferred to the lepton pair.

If the hadronic  final state is a vector meson $V$,  both the vector and axial currents contribute to the semileptonic decay $H \to V l \nu$
\bea
\langle V(p_V)| V^\mu | H(p_H) \rangle &=& V(q^2) \, \varepsilon^{\mu\sigma}_{\nu \rho} \,  \epsilon^\ast_\sigma \frac{2 p_H^\nu p_V^\rho}{m_H+m_V} \nonumber \\
\langle V(p_V)| A^\mu | H(p_H) \rangle &=& i  \epsilon^\ast_\nu \, \left[  A_0(q^2)   \frac{2 m_V q^\mu q^\nu}{q^2} \right.+
 \nonumber  \\
 &+& \left. A_1(q^2) (m_H +m_V) \eta^{\mu\nu}
- A_2(q^2)   \frac{(p_H+p_V)_\sigma q^\nu}{m_H+m_V} \eta^{\mu\sigma} \right]
 \label{LC2}
\eea
where $ \varepsilon_{\mu\sigma \nu \rho}$ is the usual totally antisymmetric tensor, $\epsilon^\mu$ is
the $D^\ast$ polarization vector  and
$\eta^{\mu\nu} \equiv  g^{\mu\nu}-q^\mu q^\nu/q^2$. Here the momentum transferred to the lepton pair is $q^\mu \equiv  p_H^\mu-p_V^\mu$, while
$m_H$ and $m_V$ are the meson $H$ and $V$ masses, respectively.

A great advantage of $B$ decays is that the mass $m_b$ of  the $ b$-quark is large compared to the QCD scale
and therefore approximations and techniques of the  heavy quark effective theory (HQET) can be used.
In  $B \to D^{(\ast)}$ semi-leptonic decays also the
mass $m_c$   of the $c$-quark can be considered  large compared to the QCD scale, allowing further approximations.
In the heavy flavour limit,
$m_{b,c} \to \infty$ ($m_b/m_c$ fixed),
when the weak current changes the flavour $b \to c$,
the light degrees of freedom inside the meson become  aware of the change in the  heavy quark velocities, $v_B  \to  v_{D^{(\ast)}} $ ($v_B \equiv p_B/m_B$, $v_{D^{(\ast)}} \equiv p_{D^{(\ast)}}/m_{D^{(\ast)}}$),
 rather than of the change in momenta.
The form factors depend on  $ \omega= v_B \cdot v_{D^{(\ast)}}$, the only scalar formed from the velocities ($v_B^2= v_{D^{(\ast)}}^2=1$ by definition). The scalar $\omega$
 is related to  $q^2$,  the momentum transferred
to the lepton pair, according to the relation
$ \omega= (m_B^2+m_{D^{(\ast)}}^2 -q^2)/(2 m_B m_{D^{(\ast)}})$.
Approximate heavy-quark symmetries
impose constraints on the form factors that
become more transparent with a  basis of form
factors different from the one given in Eqs. (\ref{LC1}) and (\ref{LC2}), that is
\bea
\frac{\langle D|V^\mu|B \rangle }{\sqrt{m_B m_D}}  &=&  h_+ (\omega) (v_B+v_D)^\mu+ h_-(\omega) (v_B-v_D)^\mu \nonumber \\
 \frac{\langle D^\ast|V^\mu|B \rangle}{\sqrt{m_B m_{D^\ast}}}  &=&  h_V (\omega) \varepsilon^{\mu \nu\rho \sigma} {v_B}_\nu {v_{D^\ast}}_\rho {\epsilon^{\ast}}_\sigma \nonumber \\
\frac{\langle D^\ast|A^\mu|B \rangle}{\sqrt{m_B m_{D^\ast}}}  &=& i h_{A_1} (\omega) (1+\omega) \epsilon^{\ast ^\mu}-i \left[  h_{A_2}(\omega) v_B^\mu + h_{A_3} (\omega) v_{D^\ast}^\mu \right] \epsilon^\star \cdot v_B
\eea
For the polarization vector,  $\sum_{\alpha=1}^3 \epsilon^{\ast \mu}_\alpha  \epsilon^{\ast \nu}_\alpha= - g^{\mu \nu} + v_{D^\ast}^\mu  v_{D^\ast}^\nu$ holds. The  factor $1/ \sqrt{m_{B(D^{(\ast)})}}$ changes the conventional, relativistic normalization  of the meson states  $ |B(D^{(\ast)})\rangle $ into a mass independent renormalization.
The form factors in the two basis are obviously related, e.g.
\beq
f_+(q^2) =  \frac{1}{2 \sqrt{r}} \left[(1+r) h_+(\omega) -(1-r) h_-(\omega) \right]
% \; f_0(q^2 )&=& \sqrt{r}\left[  \frac{\omega+ 1}{1+r}  h_+(\omega) - \frac{\omega-1}{1-r} h_-(\omega) \right]
\eeq
where $r=m_D/m_B$. Similar relations hold for the other form factors.
 Let us also define the ratios
\beq
R_1(\omega)=\frac{ h_V(\omega)}{ h_{A_1}(\omega)} \qquad R_2(\omega)= \frac{h_{A_3}(\omega)+ r h_{A_2}(\omega)}{ h_{A_1}(\omega)}\eeq that  appear in the description of $ B \to D^{(\ast)} l \nu $ decays.
In the heavy flavour limit,
there is only one form factor, the Isgur-Wise function $\xi (\omega)$ \cite{Isgur:1989vq,Isgur:1989ed}. In that limit, the form factors become
\beq
h_+ (\omega)= h_V (\omega)= h_{A_1} (\omega)  = h_{A_3} (\omega)= \xi (\omega) \qquad
h_-(\omega)= h_{A_2}(\omega)=0
\eeq

For negligible lepton masses ($l=e, \mu)$,
the  differential ratios for the semi-leptonic decays $B \to D^{(\ast)} l  \nu$  can be written as
\bea
\frac{d\Gamma}{d \omega} (B \rightarrow D\,l {\nu})  &=&  \frac{G_F^2}{48 \pi^3}\,   (m_B+m_D)^2  m_D^3 \,
(\omega^2-1)^{\frac{3}{2}}\,  |V_{cb}|^2 {\cal G}^2(\omega)\nonumber \\
\qquad\frac{d\Gamma}{d \omega}(B \rightarrow D^\ast\,l {\nu})
&=&  \frac{G_F^2}{48 \pi^3}  (m_B-m_{D^\ast})^2 m_{D^\ast}^3 \chi (\omega)  (\omega^2-1)^{\frac{1}{2}} |V_{cb}|^2  {\cal F}^2(\omega)
 \label{diffrat}
\eea
in terms of a single form factor ${\cal G}(\omega)$ and ${\cal F}(\omega)$, for $B \to D l  \nu$ and $B \to D^{\ast} l  \nu$, respectively.
 In Eq. (\ref{diffrat}),  $\chi (\omega)$  is a
 phase space factor  which reads
 \beq \chi (\omega)= (w+1)^2 \left( 1 + \frac{ 4 \omega}{\omega + 1} \frac{ m_B^2 - 2 \omega m_B m_D^\ast + m_{D^\ast}^2}{(m_B-m_D^\ast)^2} \right)\eeq
The form factor ${\cal G}(\omega)$ is a combination of $h_+(\omega)$ and $h_-(\omega)$
\beq
 {\cal G}(\omega)= h_+(\omega) -\frac{m_B-m_D}{m_B+m_D} h_-(\omega)
\eeq
Similarly, the form factor ${\cal F}(\omega)$ can be written as a function of  $h_{A_1}( \omega)$, $ R_1 ( \omega)$  and   $ R_2 ( \omega)$.
% vedi http://arxiv.org/pdf/1010.5620.pdf dopo formula (11)

The Isgur-Wise function is normalized
to unity at the zero recoil point $\omega=1$, when $D^{(\ast)}$ is at rest with respect to $B$  \cite{Shifman:1987rj, Nussinov:1986hw}.
Indeed,  at that point, the light constituents of the initial and final hadrons are not affected
by the transition $b \to c$,  and there is a complete overlap between the initial and final hadronic quantum states.
It follows
\beq
{\cal G}(1) = {\cal F}(1) =1
\eeq
Aside from short distance QCD and EW corrections, for finite values of the quark masses, the unity value of the form factors ${\cal G}$ and ${\cal F}$ is altered by  inverse powers of the masses, to be calculated nonperturbatively. Let us observe that
for ${\cal F}$ non-perturbative linear corrections are absent at zero recoil \cite{Luke:1990eg}  and the leading terms are quadratic in $1/m_{b,c}$.
We can write, schematically
\beq
{\cal F}(1) = \eta_{EW} \eta_A ( 1 + \delta_{1/m^2}+ \dots)
\eeq
where $ \delta_{1/m^2}$ are power corrections  which are suppressed by a factor of at least $\Lambda_{QCD}^2/m_c^2 \sim 3 \%$,
$\eta_{EW}$ is
the enhancement factor 1.007, due to the
electroweak corrections to the four-fermion operator mediating
the semileptonic decay \cite{Sirlin:1981ie} and $\eta_A(\alpha_s) $ is a short distance QCD coefficient known at order $\alpha_s^2$ \cite{Czarnecki:1996gu, Czarnecki:1997hc,Czarnecki:1998kt }.
A similar relation holds for ${\cal G} (1)$,  with the addition of  linear  corrections $ \delta_{1/m}$, since  in this case they, although kinematically suppressed, are not zero.

For heavy to light  transitions like $B \to \pi l \nu$, $B \to \rho l \nu$,
etc.,  the impact of  heavy quark symmetry is
less significant and is mostly reduced to flavour  symmetry relations among $B$
and $D$ decay semileptonic form factors. The form factors are generally parameterized according to Eq. (\ref{LC1}).
In the
approximation where the leptons are massless, only the form factor $f_+(q^2)$
enters the  partial rate.
In that case,
 the differential rate for, e. g.,   $ B \to \pi l  \nu$ decay reads
\beq
\frac{d \Gamma( B \rightarrow \pi l  \nu)}{dq^2}= \frac{G_F^2  |\bold{p_\pi}|^3}{24 \pi^3} |V_{ub}|^2 \, |f_+(q^2)|^2
\label{Btopiln}
\eeq
where  $\bold{p_\pi}$ is the momentum of the pion in the $B$ meson rest frame and $0< q^2< (m_B-m_\pi)^2  \simeq 26.4$ GeV. Non perturbative  theoretical predictions for form factors are usually confined to particular regions of $q^2$. Complementary regions  are spanned by
Light Cone Sum Rules (LCSR) (low $q^2$) and lattice QCD  (high $q^2$).

Lattice high-statistics calculations have been performed
 in the kinematic region where the outgoing light hadron carries little energy ($ q^2 \ge  16 \; {\mathrm{GeV}}^2$).
At low $q^2$,  with light hadrons carrying large momentum of order
2 GeV, direct
simulations require a very fine lattice which is  not yet accessible in
calculations with dynamical fermions.
The Compton wavelength of
charm and bottom quarks $\sim 1/m_{c,b}$ may be similar to or even smaller than the lattice spacing, introducing large discretization errors.  Recurring to HQET is one possible solution. By formulating the theory
such that the large energy scale is explicitly separated from the low energy
degrees of freedom, one can treat only the low energy part (which concerns the non-perturbative dynamics) in the
lattice simulation; the high energy part can be reliably treated in perturbation
theory.

The dependence of the form factor from $q^2$ is parameterized according several models, the most used ones being the  Becirevic Kaidalov (BK) \cite{Becirevic:1999kt} parameterization and  the so-called
z-expansion  \cite{Becher:2005bg, Arnesen:2005ez, Bourrely:2008za}.

LCSR combines  the standard sum rule framework with elements of the theory of hard exclusive processes.
In the sum rule approach, the $ B \ra \pi$  matrix element is
obtained from the correlation function of quark currents, such that, at large space-like external momenta, the operator-product expansion (OPE) near
the light-cone is applicable. Within OPE, the correlation function is factorized in a series
of hard-scattering amplitudes convoluted with the pion light-cone distribution amplitudes
of growing twist.
The  contributions
corresponding to higher twist and/or higher multiplicity pion  distribution amplitudes  are
suppressed by inverse powers of the $b$-quark virtuality, allowing one to
truncate the expansion after a few low twist contributions.

\subsection{Inclusive decays}
\label{subsectionInclusive decays}

Let us consider the inclusive $ \bar{B} \rightarrow X_q l \bar{\nu}$ decays, where the final state
$X_q$ is an hadronic state originated by the quark $q$. In inclusive decays, $X_q$ refers to the sum of all possible final states, no matter if single-particle  or  multi-particle states.
In the limit of large $b$-quark mass, the wavelengths associated with the $b$-quark decay
are considered  short enough to not interfere with the hadronization process.
Inclusive heavy flavour decays are regarded as occuring in two separate steps: the heavy quark decay and the final hadron composition. The second step is not expected to determine gross characteristic like total rates, etc. and quark-hadron duality is generally assumed.
Long distance dynamics of the meson can be factorized by using an  OPE approach, which, combined with HQET,
gives to inclusive transition rates  the form  of a heavy quark expansion,  schematically written as
\begin{equation}
\Gamma(B\rightarrow X_q l \nu)=\frac{G_F^2m_b^5}{192 \pi^3}
|V_{qb}|^2 \left[ c_3 \langle O_3 \rangle +
c_5\frac{ \langle O_5 \rangle }{m_b^2}+c_6\frac{ \langle O_6 \rangle }{m_b^3}+O\left(\frac{1}{m_b^4}\right)
\right] \label{HQE}
\end{equation}
Duality violation effects are hard to classify; in practice they would appear as unnaturally large coefficients of higher order terms in the expansion.
In Eq.(\ref{HQE}), $c_d$ ($d=3,5,6 \dots$) are short distance coefficients, which depend on the parton level characteristics of the hadronic final state and are calculable  in perturbation theory as a series in the strong coupling $\alpha_s$.
$O_d$ denote local operators of (scale) dimension $d$, whose hadronic
expectation values $\langle O_d \rangle $ encode the
nonperturbative corrections.
In the definition
\begin{equation}
\langle O_d \rangle  \equiv \frac{\langle B|O_d|B \rangle }{2 m_B}
\label{mb-norm}
\end{equation}
the $B$-meson mass, $m_B$, is  included for the relativistic  normalization of the state $|B>$ and for dimensional counting. These matrix elements can be systematically expanded
in powers of $1/m_b$.
A  remarkable feature  of   the total decay width in Eq.(\ref{HQE})
is the absence of a contribution of order $1/m_b$, due to the
absence of an independent gauge invariant operator of
dimension four once the equation of motion is imposed. The leading operator is $O_3 =
\bar{b}b$,  whose hadronic expectation value $ \langle \bar{b}b \rangle= 1 + O(1/m_b^2)$ incorporates the parton
model result which dominates asymptotically, i.e. for $m_b
\rightarrow \infty$.  The fact that nonperturbative, bound state
effects in inclusive decays are strongly suppressed (at least two
powers of the heavy quark mass) explains {\it a posteriori} the success
of the description in terms of the parton model.

Basically the same cast of operators in Eq.(\ref{HQE}), albeit with
different weights,  appears in semi-leptonic, radiative and
non-leptonic rates as well as distributions.
 While we can identify these operators
and their dimensions,
in general we cannot  compute their hadronic expectation
values from first principles, and we have to rely on a   number of  HQET  parameters,
which  increase with powers of  $1/m_b$.
A certain degree of universality is attained by the fact that  the
HQET parameters   not depending on the final state  appear
in different inclusive $B$ meson observables and can be measured
in experiments. By  measuring
 spectra plus as many moments as possible, one can perform
 what is generally indicated as a global fit, that is a simultaneous fit to
 HQET parameters, quark masses and absolute values of  CKM matrix elements.

In a global fit,  it is  important to understand which kind of quark
mass is to be employed,  since for confined quarks there exists no
{\it a priori } natural choice. Given a specific scheme,
masses and HQET parameters depend on it.
Using the pole masses is  computationally most convenient, but require overcoming problems due to
 misbehaved perturbative series. QCD
is not Borel summable and  the presence of (infrared) renormalons,
representing poles in the Borel plane, leads to an additive mass renormalization
generating an uncertainty in the size of the pole mass. Renormalons cancel when the inclusive decay rate is written
in terms of the  minimal subtraction ($\mathrm{\overline{MS}}$) mass, rather than the pole mass.
However, the $\mathrm{\overline{MS}}$ scheme  sets the  scale of  order of the $b$-quark mass, which is considered  unnaturally high, due to the
presence of typical scales significantly below,  down to the order of 1 GeV.
It reflects in  large corrections  at lower orders in the perturbation series.
Alternative definitions
of the  $b$-quark mass have been introduced, in order to give a better
convergence of the first (few) orders of the perturbative series and consequently reduce the
theoretical errors.

Global fits to extract $|V_{cb}|$ are currently employed  in two implementations, based on either
the kinetic scheme \cite{Bigi:1994ga,Benson:2003kp}  or the 1S scheme  \cite{Hoang:1998hm, Bauer:2004ve}. They belong to the so-called
low subtracted (or threshold) mass schemes, where non perturbative contribution to the heavy quark pole mass are subtracted by making contact to some physical observable.
Care must be taken  in converting from one mass scheme to another due to the presence of truncated
perturbative expressions.

The OPE-based, fixed order, framework just  described is not applicable in the whole phase space.
The phase space region includes a region of singularity, also called endpoint or threshold region,
 corresponding to a kinematic region near the limits of
both the lepton energy  $E_l$ and $q^2$ phase space, where the rate is dominated by
the production of low mass final hadronic states.
This region is
plagued by the presence
 of large double (Sudakov-like)  perturbative  logarithms at all orders in the strong coupling.
Corrections  can be large  and need to be resummed at all orders.
This can be intuitively understood, since  by integrating
out the heavy flavor masses in HQET, the
only remaining scales in the hadronic subprocess
are $m_X$ and a hard scale $E_X$, where $m_X$
is the invariant mass of the hadronic system $X_q$ and $E_X$ its energy. Infrared logarithms
occur in the ratio $E_X/m_X$ and become large in the threshold region, where $m^2_X \sim O(E_X \Lambda_{\mathrm{QCD}})$.
A resummation formalism  analogous to the one used to factorize Sudakov threshold effects for
parton distribution functions in usual hard processes, such as deep inelastic
 or Drell-Yan scattering, can be applied\footnote{for
 theoretical aspects of threshold
resummation in $B$ decays
see Refs.~ \refcite{DiGiustino:2011jn,Aglietti:2007bp,Aglietti:2005eq,Aglietti:2005bm}, \refcite{Aglietti:2005mb}, \refcite{Aglietti:2002ew},\refcite{Aglietti:2000te}, \refcite{Aglietti:1999ur}, and references therein.}.
The relative low energy threshold region is also sensitive to non perturbative effects.
One is
the so-called   Fermi motion, which classically can be described as a
small vibration of the heavy quark inside the B meson due to the momentum exchange with the valence
quark. This effect is important
in the end-point region, because it produces some smearing of the partonic spectra. Let us illustrate it with
the simplest example,  occurring in the inclusive radiative decay $B \to X_s \gamma$.
To leading order in $ 1/ m_b$,  the  $B \to X_s \gamma$ decay is described
by the quark level transition $ b \to s \gamma$ and no events can be generated beyond the quark level kinematical boundary,
i.e. with $E_\gamma > m_b/2$. On the other hand, the true kinematical boundary is set by
the higher hadron mass $m_B/2$.
It is intuitively clear the physical solution: the $b$-quark is not at rest inside the B meson and its Fermi motion spreads
the photon line out over a region of order $m_B-m_b$.

For  $b \rightarrow c$ semileptonic decays, the effect of the small region of singularity is not very important; in addition,  corrections are not expected  as singular as in the $ b \rightarrow u$ case, being  cutoff by the charm mass.
However, in $ B \to X_u l \nu$
decays, the copious background from the $ B \to X_c l \nu$ process, which has a rate about 50 times higher,
stands in the way, and data taken is pushed towards restricted regions of phase space,
where such background is highly suppressed by kinematics.
The OPE framework can reliably predict the inclusive  decay rate
as long as it is integrated over a large region of the phase space. The experimental
cuts required to suppress the  background violate this requirement.
The weight of  the threshold region within the phase space region increases and
 the above mentioned   theoretical issues need to be addressed.
Several theoretical approaches have been proposed, that will be discussed in Sect. \ref{vuninclusivo}.

Finally, let us observe that,
very recently, the branching fractions of  $B_s \to X l \nu$ decays have been measured  at BaBar \cite{Lees:2011ji} and Belle \cite{Oswald:2012yx},
in  datasets obtained from  energy scans above the $\Upsilon(4S)$, with uncertainty going down as much as  5-6 \% \cite{Oswald:2012yx}.
In this section, we have always implicitly alluded to $B$ decays, but  semileptonic $B_s$ decays can also  probe CKM matrix elements, within the approach just outlined. The  presence of the heavier spectator strange quark is bound to introduce  some amount of SU(3) symmetry
breaking.

\section{$|V_{cb}|$ }
\label{vcb}

In 1983, Mark II and MAC collaborations presented the first measurements of the average lifetime of $b$-quark, obtained at the $e^+ e^-$ storage ring  PEP  located at SLAC,   and  gave the first estimate of $|V_{cb}|$ (setting  $|V_{ub}|$ to zero) \cite{Lockyer:1983ev, Fernandez:1983az}.
 Currently, $|V_{cb}|$ is  inferred from semi-leptonic decays;
about 25\% of all B mesons decay semileptonically via the tree-level $ b \to c$  quark
transition.

\begin{figure*}[ht]
\centerline{\includegraphics[width=2.9in]{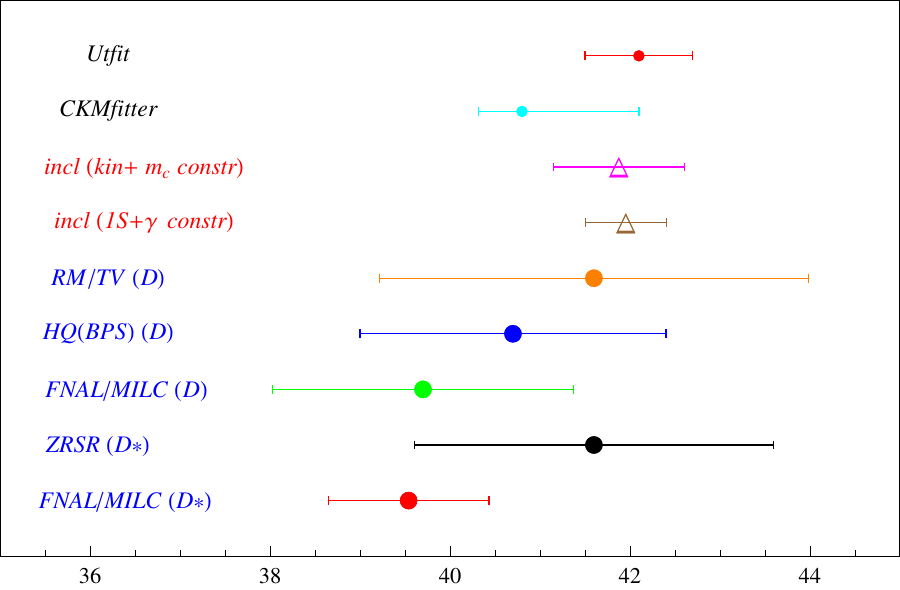}}
\vspace*{29pt}
\caption{Comparison among exclusive, inclusive and indirect determinations of $|V_{cb}|$ ($10^{-3}$).
For the exclusive decays the legend refers to
the theoretical determination of the form factor: for $B \to D^\ast l \nu $, unquenched  lattice FNAL/MILC \cite{Bailey:2010gb} and ZRSR  \cite{Gambino:2010bp, Gambino:2012rd};
for $B \to D  l \nu $, unquenched  lattice FNAL/MILC \cite{Laiho:2005ue}, quenched lattice Rome/Tor Vergata (RM/TV) \cite{de Divitiis:2007ui, deDivitiis:2007uk}, HQ/BPS expansion \cite{Uraltsev:2003ye}. For inclusive decays, HFAG \cite{Amhis:2012bh} results are reported in the kinetic scheme with the $m_c$ constraint, and in the 1S scheme with the $ B \to X_s \gamma$ constraint. Indirect fits are from Utfit \cite{Silvestrini} and CKMfitter collaborations \cite{CKMfitter}.
\protect
\label{Vcb-fig}}
\end{figure*}

In Fig. \ref{Vcb-fig} we plot recent $|V_{cb}|$ determinations, exclusive and inclusive, discussed in detail in Sects. \ref{vcbexcl1} and \ref{incldec1sect}. To facilitate the comparison,  the errors shown are the squared average of theoretical and experimental errors.
We can observe  a certain tension between the inclusive  and the exclusive unquenched lattice results, around $ 2\sigma$ with the most recent lattice results by FNAL/MILC \cite{Bailey:2010gb}, having a considerable reduced theoretical error. New lattice results  are in progress and expected soon.
We also compare with
$ |V_{cb}|  = (42.07 \pm  0.64) \times 10^{-3}$ by the UTfit collaboration \cite{Silvestrini} and
$ |V_{cb}|  = (40.77^{+0.13}_{-0.48}) \times 10^{-3}$ by the CKMfitter collaboration (at $1 \sigma$) \cite{CKMfitter}.
Indirect fits  prefer a value for $|V_{cb}|$ that is closer to the (higher)
inclusive determination.

%\begin{figure*}[ph]

\subsection{Exclusive determination}
\label{vcbexcl1}

 Let us consider  the tree level driven
$ \bar{B}\rightarrow D^{(\ast)}  l \bar{\nu}$
weak decays, where $l$ is an electron or a muon.
Neglecting  the charged lepton and neutrino masses,   their differential ratio  can be parameterized in terms of  the form factors ${\cal G (\omega)}$ and ${\cal F}  (\omega)$, according to Eq. (\ref{diffrat}).

The extraction of $|V_{cb}|$ can be reckoned as divided into two steps, the first one being the experimental fit of  the products
$|{\cal G}(\omega) V_{cb}| $ and $|{\cal F}(\omega) V_{cb}|$.
 Due
 to the kinematic suppression
factors, $(\omega^2-1)^{3/2}$ and $(\omega^2-1)^{1/2}$, data are taken at $\omega
\neq 1$.
Step two is the theoretical evaluation of the form factors.
It is generally performed at zero-recoil point, where
the non perturbative evaluation of the operator matrix elements is simplified by  heavy flavour approximations and symmetries.
Since the zero recoil point is not accessible experimentally, the $|V_{cb}|$ estimates rely on the
extrapolation from $\omega \neq 0$ to the zero recoil point.
The dynamics of the decay is contained in the form factors
${\cal F}(\omega)$ and ${\cal G}(\omega)$,  which can be parameterized
in  model dependent ways based on the HQET framework \cite{Caprini:1997mu, Boyd:1995sq, LeYaouanc:2002vp}.
The experimental fit of step one
 may also extend to parameters characterizing the  dependence  on $\omega$  of the form factors.

For the determination of  $|V_{cb}|$, the decay $ \bar B \to D^\ast l \bar \nu $ is preferred over
 $\bar  B \to D l \bar \nu $ for both theoretical and experimental reasons.
At present, the $ \bar{B}\rightarrow D^\ast  l  \bar{\nu}$   decay, with an higher rate,  is measured with a better
accuracy than the $ \bar{B}\rightarrow D l  \bar{\nu}$  decay.
On the theory side,  nonperturbative  linear corrections to the form factor  ${\cal F}$ are absent at zero recoil.
 The most recent  Heavy Flavor Averaging
Group  (HFAG) experimental  fit \cite{Amhis:2012bh} gives
\beq  |V_{cb}| |{\cal F}(1)| = (35.90 \pm 0.45 ) \times   10^{-3} \label{VcbexpF1} \eeq
%
%
%
% Vedi http://arxiv.org/pdf/hep-ex/0602023.pdf
%
The currently most precise measurement  \cite{Dungel:2010uk} uses 711 ${\mathrm{fb}}^{-1 }$ of data collected by the Belle experiment  and  fits  four kinematic variables
fully characterizing the form factor in the  framework of Ref. \refcite{Caprini:1997mu}, that is
$F(1)|V_ {cb}|$,  $R_1(1)$, $R_2(1)$ and  a parameter $ \rho^2$.
Their results agree well with the ones from BaBar \cite{Aubert:2007rs}.

%More precision than in general form factor evaluation are in lattice evaluation for semileptonic  since possible to connect to ratio
%(or double ratios)  where most uncertainties cancel .

Nonperturbative corrections to the unit limit of   $ {\cal F}(1) $ can be theoretically computed by means of
lattice QCD.
%
% Other solutions may require the use of different effective theories, e.g. NRQCD, or  relativistic (improved) formulations.
%
The first lattice calculation for  $ {\cal F}(1) $ has been accomplished by FNAL \cite{Hashimoto:2001nb} in the quenched approximation, in which the effect of vacuum
polarization of quark loops is neglected.
Quenched lattice results are also available at finite momentum transfer ($\omega= 1.075$) \cite{deDivitiis:2008df}, and combined with 2008 BaBar data \cite{Aubert:2007rs} give
\beq
|V_{cb}| =  (37.4 \pm 0.5_{\mathrm{exp}} \pm 0.8_{\mathrm{th}} ) \times 10^{-3}
\label{latticeTV}
\eeq
with a rather small nominal error.

Unquenched calculations take  into account vacuum polarization effects, i.e., include up, down and
strange sea quarks on the gauge configurations' generation. The up and down quarks are usually
taken to be degenerate, so those simulations are referred to as $n_f= 2+1$.
The only unquenched calculations available for ${\cal F}(1)$  have been performed   by FNAL/MILC  \cite{Bernard:2008dn, Bailey:2010gb}.
Their  latest  update \cite{Bailey:2010gb} reduces the total uncertainty on $ {\cal F}(1)$ from about  2.6\%  to 1.8\% and gives
\beq  {\cal F}(1)
=0.908\pm 0.017   \label{VcbexpF2}  \eeq
by using  lattice with  the Fermilab action for $b$- and $c$-quarks, the asqtad
staggered action for light valence quarks, and the MILC ensembles for gluons and light quarks.
It includes the enhancement factor 1.007, due to the
electroweak  (EW) corrections to the four-fermion operator mediating
the semileptonic decay \cite{Sirlin:1981ie}.
By combining Eq. (\ref{VcbexpF2}) with the HFAG results in Eq. (\ref{VcbexpF1}),
one estimates
\beq
|V_{cb}| =  (39.54 \pm 0.50_{\mathrm{exp}} \pm 0.74_{\mathrm{th}} ) \times 10^{-3}
\label{lattice53}
\eeq
Let us observe that  a further  update  by  FNAL/MILC collaboration  has  been announced   \cite{JackLaihotalk}, claiming a reduction of discretization effects and of the error on $|V_{cb}|$ down to 1.6\%, but no new value for $|V_{cb}|$ has been published until now.

The lattice calculations have to be compared with non-lattice ones.
By  using zero recoil sum rules,
 the value
\beq {\cal F}(1) = 0.86 \pm 0.02 \label{gmu} \eeq
 has  been  recently reported  \cite{Gambino:2010bp, Gambino:2012rd}.
Let us remark that  full $\alpha_s$ and up to $1/m^2$ corrections are included in this result; morever,  in order to compare  with the lattice value in Eq.~(\ref{VcbexpF2}),
one has to remove the EW factor 1.007 from the latter.
The related estimate, given the HFAG average in Eq.~(\ref{VcbexpF1}), yields to
\beq
|V_{cb}| =  (41.6\pm 0.6_{\mathrm{exp}}\pm 1.9_{\mathrm{th}}) \times 10^{-3}
\label{VCBF1}
\eeq

The theoretical error is more than twice the error in the lattice determination given in Eq. (\ref{lattice53}).
However, let us observe that
the budget error from lattice has been recently questioned~\cite{Gambino:2012rd}. The claim is that existing  differences between the power-suppressed deviations from the heavy flavour  symmetry
 in
the lattice theory with heavy
quarks and in continuum QCD may be compensated by a matching between the two theories that has  been performed, at the best, only at
lower levels.

Let us compare the previous determinations with the $|V_{cb}| $ value extracted from  $ \bar{B}\rightarrow D \, l \, \bar{\nu}$ decays.
The HFAG  average   includes older Aleph, CLEO and Belle measurements, as well as the new 2008-2009 BaBar  data,
and adopts  the parametrization in Ref. \refcite{Caprini:1997mu}, where the
form factor ${\cal G}(\omega) $  is described by only two parameters: the normalization ${\cal G}(1) |V_{cb}|$ and the slope $\rho^2$.
The resulting global two-dimensional fit gives \cite{Amhis:2012bh}
\beq |V_{cb}| | {\cal G}(1)| = (42.64 \pm 1.53 )  \times
10^{-3} \label{BDfitdat} \eeq
Unquenched calculations of the  form factor $ {\cal G}(1) $  have been performed by the FNAL/MILC collaboration in 2005
\cite{Okamoto:2004xg}, with an update the year after \cite{Laiho:2005ue},
giving \beq {\cal G}(1) = 1.074 \pm 0.024 \label{latBD} \eeq
after
 correcting by the usual EW factor of 1.007.
Let us observe that studies  of this form factor  at non-zero recoil are in progress and  the FNAL/MILC collaboration has already reported some preliminary results \cite{Qiu:2012xi}.
By combining Eqs. (\ref{BDfitdat}) and  (\ref{latBD}),
the resulting  estimate is
 \beq |V_{cb}| = (39.70 \pm 1.42_{\mathrm{exp}} \pm 0.89_{\mathrm{th}})
\times  10^{-3} \label{lattice2} \eeq
 in good agreement with the lattice determination from  $ \bar{B}\rightarrow D^\ast l  \bar{\nu}$,  Eq. (\ref{lattice53}), although
the experimental error  is more than twice larger.
At non-zero recoil,   a lattice determination is already available, but only in the quenched approximation  \cite{de Divitiis:2007ui, deDivitiis:2007uk}. By
 using 2009  BaBar  data \cite{Aubert:2009ac}, a slightly higher value is found
 \beq |V_{cb}| = (41.6 \pm 1.8 \pm 1.4
\pm 0.7_{FF} ) \times  10^{-3} \label{lattunq1} \eeq
The errors are  statistical, systematic and due to the theoretical uncertainty in the form factor $ {\cal G}$, respectively.

The most recent non lattice calculation dates 2004 and  combines the heavy quark expansion
with expanding around the point
where the kinetic energy
 is equal to
the chromomagnetic moment
 $\mu_\pi^2=\mu^2_G$ (''BPS" limit) \cite{Uraltsev:2003ye}.
A number of relations, connected to the form factor, receive corrections
only to the second order expanding around this limit to any order in $1/m_{c,b}$.
%
%It has been also motivated by  rather close values of these two parameters obtained  from experiments in  inclusive B decay.
%
Under this approximations, the  form factor reads
 \beq {\cal G}(1) =1.04 \pm  0.02 \eeq
With such estimate the PDG finds \cite{Beringer:1900zz}
\beq
|V_{cb}| =  (40.7 \pm 1.5_{\mathrm{exp}} \pm 0.8_{\mathrm{th}}) \times 10^{-3}
\eeq
in agreement, within the errors, with both lattice
determinations  (\ref{lattice2})  and (\ref{lattunq1}).

Semileptonic B decays to orbitally-excited P-wave
charm mesons ($ D^{\ast \ast}$)
contribute as a source of systematic error in the $|V_{cb}|$ measurements at the B factories (and
previously at LEP), as a background to the direct decay $ B^0 \to D^{\ast } l  \nu$ (see BaBar preliminary results in Ref. \refcite{Margoni:2013qx}).
The  knowledge on these semileptonic
decays is not complete yet: one example for all,  the so called "1/2 versus 3/2" puzzle\footnote{for a brief  status report   on $B$ semileptonic decays to excited $D$ states see e.g. Sect. 3 of Ref.  \refcite{Ricciardi:2012pf}.}.
Very recently, first dynamical lattice computation of the $\bar B \to D^{\ast \ast } l \nu$ form factors have been attempted, although  still preliminary and needing extrapolation to the continuum \cite{Atoui:2013sca}.

Semileptonic decays of B mesons to the $\tau$  lepton are experimentally  challenging to
study because their rate is  suppressed, due to  the large $\tau$ mass, and
the final state contains not just one, but
two or three neutrinos as a result of the $\tau$ decay.
Theoretically, an additional form factor is needed for both $\bar B \to D \tau^- \bar \nu_\tau$ 
and $\bar B \to D^\ast \tau^- \bar \nu_\tau$, since the $\tau$ mass cannot be neglected. 
The first exclusive observation of $ B^0 \to D^{\ast -} \tau^+ \nu_\tau$ decays was presented by Belle in 2007 \cite{Matyja:2007kt}.
Since then, both BaBar and Belle have published improved measurements, and
have found evidence for $ \bar B \to D \tau^- \bar \nu_\tau$ decays. 
The branching ratio measured values have
consistently exceeded the SM expectations and the experimental precision starts to be enough to constrain NP.
The most recent data from BaBar are not compatible with a
charged Higgs boson in the type II two-Higgs-doublet model
and with large portions of the more general type III two-Higgs-doublet model \cite{Lees:2013uzd}.
At present, semileptonic  $b \to \tau$  decays
do  not  contribute to the determination
of $|V_{cb}|$,  but are studied because of their NP sensitivity.
The same is true for
 exclusive $B_s$ decays, that  are attracting a lot of attention, due to the avalanche of recent data
and to the expectation of new ones\footnote{see, e.g., Refs. \refcite{Ricciardi:2012pf}, \refcite{Ricciardi:2012qm} and references therein.}.

\subsection{Inclusive decays}
\label{incldec1sect}

In Sect.\ref{subsectionInclusive decays}, the standard setting  for the $|V_{cb}|$ extraction  using data on  inclusive semileptonic decays $B \to X_c l \nu$ has been outlined.
The expansion (\ref{HQE}) is  valid only for sufficiently inclusive measurements
and away from perturbative singularities, therefore the relevant quantities to be measured are
global shape parameters (the first few moments of various kinematic distributions)
and the total rate.
 As already discussed in Sect.\ref{subsectionInclusive decays}, masses and HQET parameters need to be defined in a given mass scheme;
 global fits to extract $|V_{cb}|$, currently employed by HFAG \cite{Amhis:2012bh},  have been performed in two schemes, the kinetic and the 1S schemes.
Care must be exercised while  passing from one scheme to another, due to the presence of
 different definitions and  approximations.
The reliability of the inclusive method depends also on the ability to control the higher order
contributions in the expansion (\ref{HQE}), a  double series in $\alpha_s$ and $\Lambda_{\mathrm{QCD}}/m_{b}$.
 The calculation of higher order effects permits to ascertain unwanted behavior
of the double series and to reduce  the theoretical uncertainty due to the truncation.

The  leading term is the parton model,
which is known completely to order $\alpha_s$ and $\alpha_s^2$,
for the width and moments of the
lepton energy and hadronic mass distributions
 (see Refs.  \refcite{Aquila:2005hq}, \refcite{Pak:2008qt}, \refcite{Pak:2008cp}, \refcite{Biswas:2009rb}
and references therein).

The coefficients of
even the first non-perturbative corrections are not completely known to order $\alpha_s$.
In the expansion,  the leading power corrections arise from
two  dimension five operators: the kinetic operator  $  O_{kin}$  and the chromomagnetic operator $O_{mag}$.
Different schemes are used to define these parameters, but to leading order and
leading power they are given by
\bea
\langle O_{kin}\rangle &\equiv&  \frac{1}{2 m_B} \langle \bar B(p_B)| \bar b_v (i D)^2 b_v | \bar B(p_B)\rangle \equiv  - \mu_\pi^2 \nonumber \\
\langle O_{mag}\rangle &\equiv&  \frac{1}{2 m_B} \langle \bar B(p_B)| \bar b_v \frac{g}{2} \sigma_{\mu\nu} G^{\mu\nu} b_v  | \bar B(p_B)\rangle \equiv  \mu_G^2 \label{statesHQ}
\eea
where $b_v$ is the quark field in the HQET and $D$ is the covariant derivative with respect to the background
gluon field. Sometimes in the definitions (\ref{statesHQ})  the spatial component $D_\perp^\mu= (g^{\mu\nu}-v^\mu v^\nu) D_\nu$ is used instead;  these expressions differ by higher-order terms in
the expansion $1/m_b$.
The HQET parameters $\mu_\pi^2$ and $\mu_G^2$ are also
 denoted by $-\lambda_1$ and  $ 3 \lambda_2$, respectively \cite{Falk:1992wt}.
 For the
total rate,  the kinetic corrections have the same coefficient as the leading order.
For other observables, such as partial rates and moments,
the corrections to the coefficient of the  kinetic matrix element  have been
 evaluated   at $O(\alpha_s)$ order \cite{Becher:2007tk, Alberti:2012dn}.
They lead to
numerically modest modifications of the width and moments.
Corrections at order $O(\alpha_s$) to the
coefficient of  the matrix element of the chromomagnetic operator are not yet available, although a study is in progress
 \cite{Alberti:2012dn}.
In the
simpler case of inclusive radiative decay, these corrections
have increased the
coefficient by almost 20\% in
the rate \cite{Ewerth:2009yr}.

At order $1/m_b^3$ the expansion (\ref{HQE}) receives contributions
from  local dimension-six operators. There  are also other sources of
$1/m_b^3$
corrections.
The 
matrix elements of Eq. (\ref{statesHQ}) have an implicit dependence on $m_b$.
At order
 $1/m^2_b$,
this dependence can   be neglected, but at higher orders this mass dependence has to be taken
into account explicitly.
Neglecting  perturbative corrections, i.e. working at tree level,  contributions to various observables   have been
computed to order $1/m_b^4$ \cite{Gremm:1996df, Dassinger:2006md}.
At order $1/m^3_b$,
terms with a
 sensitivity
to the charm mass $m_c$ start to acquire relevance \cite{Bigi:2005bh, Breidenbach:2008ua, Bigi:2009ym}.
Roughly speaking, since $m^2_c \sim O( m_b \Lambda_{\mathrm{QCD}})$ and $\alpha_s(m_c) \sim O(\Lambda_{\mathrm{QCD}})$, contributions of order
 $1/m^3_b \, m^2_c$
and $\alpha_s(m_c)1/m^3_b\, m_c
$  are expected
comparable in size to the contributions of order $1/m^4_b$.
Contributions  $O(1/m^{4,5}_b)$ have been
estimated in the ground state saturation approximation  \cite{Mannel:2010wj}, resulting in a small   0.4\% increase of $|V_{cb}|$.
The usefulness of the ground state saturation has
been recently questioned \cite{Gambino:2012rd}, on the basis that
the non-factorizable contributions can
in general be comparable to the factorizable ones.

In order to perform a global fit to  $|V_{cb}|$, the $ b$-quark mass and the
 hadronic parameters,
HFAG  employs as  experimental inputs  the (truncated) moments of the
lepton energy $E_l$  (in the $B$ rest frame) and the $m_X^2$  spectra in $B \to X_c l \nu$ \cite{Amhis:2012bh}.
 A total of about 70 measurements is available,  80\% of which  performed at the $B$-factories.
Let us underline that, since
new non-perturbative parameters
appear  at each order in $1/m_b$ (e.g. as many as nine new expectation values  at
$O(1/m^4_b )$), only the parameters associated with  $O(1/m^{2,3}_b)$  corrections are
routinely fitted from experiment.

 The moments in $ B \to
X_c l \nu$ are sufficient
for determining  $|V_{cb}|$, but measure the $b$-quark mass only to about 50 MeV precision. To get higher precision,
additional constraints are introduced:  the photon energy moments in
 $ B \to
X_s \gamma$,  or a precise  constraint  on the $c$-quark  mass.
Using the  former constraint, in the kinetic scheme,
the global fit yields
  \beq |V_{cb}| = (41.88 \pm 0.73) \times 10^{-3}\eeq
with the value
$m_c^{\overline{\mathrm{MS}}} \mathrm{(3 GeV)} = (0.998 \pm 0.029)$ GeV, obtained using low-energy sum rules \cite{Dehnadi:2011gc}.
In the 1S scheme, the $c$-quark mass
constraint cannot be applied as the 1S expressions
do not depend on this parameter. The result, using the
 $ B \to
X_s \gamma$ constraints,
is
\beq |V_{cb}| = (41.96 \pm 0.45) \times 10^{-3}\eeq
The central values are in excellent agreement  in the two schemes.
The precision is
higher than in the exclusive determinations, being  about 1.7\%
in the kinetic scheme and  1.1\% in the 1S scheme.

High statistic $B$-factories have greatly contributed to the increase in measurement precision with respect to previous experiments.
BaBar and Belle have collected about 1.5 ab$^{-1}$ high-quality data, and Belle II at SuperKEKB is expected to collect about 30 times more data by 2023, pushing the error on $|V_{cb}|$ down to 1\% \cite{Yashchenko}.

%,due to a more aggressive
%error estimate.
%(V Scwhanda, che prende i dati da HFAG 2012 1207.1158 pag 129)

\section{$|V_{ub}|$}
\label{vub}

\begin{figure*}[th]
\centerline{\includegraphics[width=2.9in]{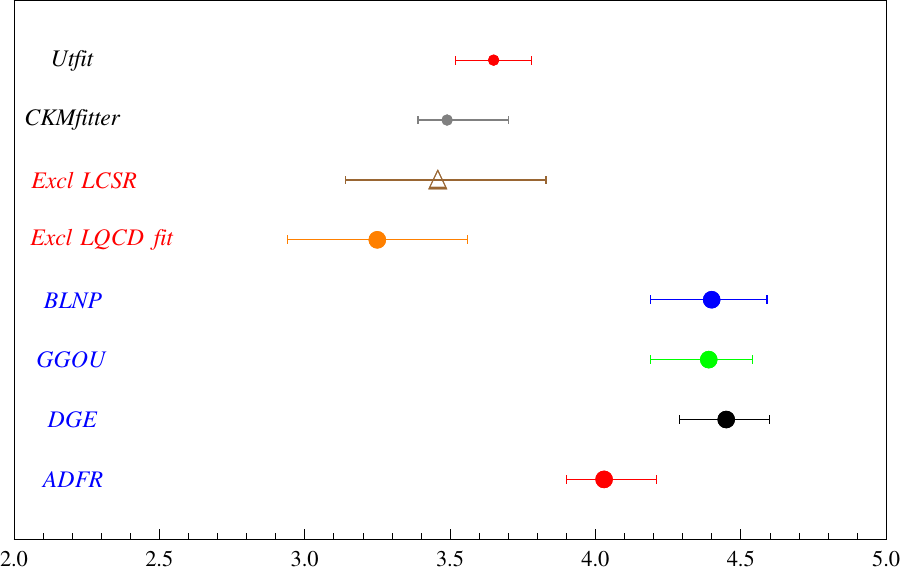}}
\vspace*{29pt}
\caption{Comparison among exclusive, inclusive and indirect determinations of $|V_{ub}|$ ($10^{-3}$).
 For inclusive decays,  we refer to the HFAG estimates
\cite{Amhis:2012bh},
following ADFR \cite{Aglietti:2004fz, Aglietti:2006yb,Aglietti:2007ik}, DGE \cite{Andersen:2005mj}, GGOU \cite{Gambino:2007rp}, BLNP \cite{Lange:2005yw, Bosch:2004th, Bosch:2004cb}
determinations. For the exclusive decays we report, from the most recent estimates \cite{Lees:2012vv},  the full $q^2$ range fit to  data and FNAL/MILC LQCD results  \cite{Bailey:2008wp} and the QCDSR based  determination   below $q^2 = 16 $ GeV  \cite{Khodjamirian:2011ub}. Indirect fits are from Utfit \cite{Silvestrini} and CKMfitter collaborations \cite{CKMfitter}.
The largest  among theoretical and experimental errors  have been depicted.
\protect
\label{Vub-fig}}
\end{figure*}

It is likely the most studied CKM matrix element, on both  theoretical and experimental aspects, but it is, comparatively,  the less known.  The order of magnitude of the ratio $|V_{ub}/V_{cb}|\sim 0.1$ has been known since the ' 90s \cite{Albrecht:1989qv,Fulton:1989pk}. The  error on $|V_{ub}|$,  that at the time was around $30\%$, is  now reduced of about 1/3. The increased precision has made manifest
a tension between the values of $|V_{ub}|$ extracted from the exclusive and inclusive semileptonic decays.
In Fig.
\ref{Vub-fig} we show some recent exclusive and inclusive determinations,  that will be commented in Sects. \ref{Exclusivesemileptonicdecays33},  \ref{Exclusivesemileptonicdecays34} and
\ref{vuninclusivo}.

%\begin{figure*}[ph]
The  leptonic decay $ B \to \tau \nu$, first  observed  by Belle in 2006 \cite{Ikado:2006un}, can also provide information on $|V_{ub}|$, which we do not display in Fig.
\ref{Vub-fig}.
Indeed,
previous data have shown a disagreement of the measured branching ratio with the SM prediction, which  has softened significantly with the new data from Belle
collaboration \cite{Adachi:2012mm}.

%$|V_{ub}|$ is connected to CP violation observables;
% the compatibility of $\sin 2 \beta $ from $B \to J/\psi K$ decays strongly depends on input from $|V_{ub}|$. (vedi silvestrini and Utfit; anche Buras e Soni vecchi talks)

In Fig.
\ref{Vub-fig}, we also compare with indirect fits,
$ |V_{ub}|  = (3.65 \pm  0.13) \times 10^{-3}$ by  UTfit \cite{Silvestrini} and
$ |V_{ub}|  = (3.49^{+0.21}_{-0.10}) \times 10^{-3}$ at $1 \sigma$ by CKMfitter \cite{CKMfitter}.
At variance
with the $|V_{cb}|$ case, the results of the global fit prefer a value for $|V_{ub}|$ that is closer to the (lower)
exclusive  determination.

%(V anche http://arxiv.org/pdf/1303.3723.pdf pag 28)

%http://arxiv.org/pdf/1211.1896.pdf  pag 29

%forse i primi vanno come input

\subsection{Exclusive semileptonic decays}
\label{Exclusivesemileptonicdecays33}

Among all charmless $B$ meson semileptonic channel presently observed,
  $ B \rightarrow \pi l  \nu$ decays benefit of more precise branching fraction measurements  and are currently the  channels of election
 to
determine  $|V_{ub}|$ exclusively.
The  $ B \rightarrow \pi l  \nu$  decays are affected by a single form factor $f_+(q^2)$, in the limit of zero leptonic masses (see the 
differential ratio in Eq. (\ref{Btopiln})).
%
%By using current lattice
%QCD methods, the hadronic
%amplitudes for  $\bar B \rightarrow \pi l \bar \nu_l$  can be calculated quite accurately  because  there is only a single %stable
%hadron in both the initial and final states.
The  first lattice determinations of  $f_+(q^2)$  based on unquenched  simulations have been obtained by the Fermilab/MILC collaboration\cite{Bailey:2008wp}  and the HPQCD collaboration
\cite{Dalgic:2006dt}, and they are  in substantial agreement.
In Ref. \refcite{Bailey:2008wp},  the $b$-quark
is simulated by  using  the so-called Fermilab heavy-quark method, while
 the dependence of the form factor from $q^2$ is parameterized according to the
z-expansion  \cite{Becher:2005bg, Arnesen:2005ez, Bourrely:2008za}.
In Ref. \refcite{Dalgic:2006dt},
the $b$-quark
is simulated by using nonrelativistic QCD and  the BK  parameterization \cite{Becirevic:1999kt}  is  extensively used for the $q^2$ dependence.
Recent results are also available on a fine lattice (lattice spacing $a \sim 0.04$ fm) in the quenched approximations by the QCDSF collaboration \cite{AlHaydari:2009zr}.

Latest data  on   $ B \rightarrow \pi l  \nu$ decays coming from Belle and BaBar \cite{Bailey:2008wp, Lees:2012vv} are  not yet included in the HFAG averages \cite{Amhis:2012bh}.
The measured
partial branching fractions can be fit at low and high $q^2$  according to LCSR and lattice approaches, respectively, the latter providing generally better fits.
Both BaBar and Belle  collaborations  determine the  magnitude of the CKM matrix element $|V_{ub}|$
 using two different methods.
In one case, the $|V_{ub}|$ value is extracted in a limited range of $q^2$
from the measured
partial branching fraction using the relation $ |V_{ub}| =
\sqrt{ \Delta {\cal{ B}} /(\tau \Delta \zeta)} $, where $\tau$  is the $B$  lifetime  and
$\Delta \zeta = \Gamma /|V_{ub}|^2$ is the normalized partial decay width derived in different
theoretical approaches.
In the other,  a simultaneous
fit to  lattice results  and
experimental data is performed, to exploit all the available
information on the form factors from the data (shape) and
theory (shape and normalization).
The simultaneous fit to the data over the full $q^2$ range and the FNAL/MILC
 lattice QCD results  \cite{Bailey:2008wp} performed by  Belle \cite{Ha:2010rf} has
given  the following average value
\beq |V_{ub}|  = (3.43 \pm  0.33) \times 10^{-3}\eeq
More recently, the analogous fit by BaBar \cite{Lees:2012vv} has yielded
\beq
|V_{ub}| = (3.25 \pm 0.31) \times 10^{-3} \label{exclus}\eeq
in the full $q^2$ range.
The $|V_{ub}|$ determinations inferred
by  using lattice QCD
 direct calculations,
 in the kinematic region $q^2 > 16$ GeV$^2$, with no extrapolations,
are in agreement with the previous values \cite{Bailey:2008wp, Lees:2012vv}.

In the complementary   kinematic region, at
large recoil,  with an upper limit for $q^2$ varying between 6 and 16 GeV$^2$,
several  direct calculations of the semileptonic  form factor  are available,   based on LCSR. There has been recent progress in pion distribution amplitudes, NLO and LO higher order twists (see e.g.~ \refcite{Khodjamirian:2011ub},\refcite{Ball:2001fp},\refcite{Ball:2004ye},\refcite{Ball:2007hb},\refcite{Khodjamirian:2000ds},
\refcite{Duplancic:2008ix}, \refcite{Bharucha:2012wy}, \refcite{Li:2012gr} and references therein).
The latest determination, from BaBar collaboration \cite{Lees:2012vv},
 using  LCSR  results   below $q^2 = 16 $ GeV  \cite{Khodjamirian:2011ub}, gives
\beq |V_{ub}| =
(3.46 \pm 0.06 \pm 0.08^{+0.37}_
{-0.32})  \times 10^{-3}  \eeq
 where the  three uncertainties  are statistical, systematic and theoretical, respectively.
The Belle collaboration \cite{Ha:2010rf} has  used the LCSR   form factor determination in Ref. \refcite{Ball:2004ye}, with the same $q^2 = 16 $ GeV as upper limit,  and found a consistent value 
 \beq  |V_{ub}|  =
(3.64 \pm 0.11^{+0.60}_
{-0.40})  \times 10^{-3}  \eeq

% The latest   lattice average  gives \cite{elk}  \beq
%|V_{ub}|  = (3.12 \pm 0.26)\times 10^{-3} \eeq

Recently, BaBar and Belle collaborations have reported significantly improved branching ratios of other
 heavy-to-light semileptonic decays, that reflects on
increased precision
for $|V_{ub}|$ values inferred by these decays.
$|V_{ub}|$ has been extracted from  $ B^+ \rightarrow \omega l^+ \nu$, with the LCSR form factor determination \cite{Ball:2004rg} at $q^2 < 20.2$ GeV, yielding  \cite{Lees:2012vv}
\beq
|V_{ub}|  =
(3.20 \pm 0.21 \pm 0.12^{+0.45}_
{-0.32})  \times 10^{-3}  \eeq
where the three uncertainties
are statistical, systematic and theoretical, respectively.
By comparing the
measured distribution in $q^2$, with an upper limit at $q^2 = 16$ GeV, for  $ B \to \rho l \nu$ decays,
with LCSR  predictions for the form factors\cite{Ball:2004rg},
 the $|V_{ub}|$ value reads  \cite{delAmoSanchez:2010af}
\beq
|V_{ub}|  =
(2.75 \pm 0.24 ) \times 10^{-3}  \eeq
Other interesting  channels  are
 $ B \rightarrow \eta^{(\prime)} l \nu $ \cite{delAmoSanchez:2010zd, DiDonato:2011kr}, but  a value of
$|V_{ub}|$ has not been extracted  because
the theoretical partial decay rate is not sufficiently
precise yet.
There has also been    recent progress   on the form factor evaluation of the $|V_{ub}|$ sensitive $\Lambda_b \to p l \nu$ decay  in the LCSR framework
\cite{Khodjamirian:2011jp} and from lattice with static $b$-quarks \cite{Detmold:2013nia}.

\subsection{Purely leptonic decays}
\label{Exclusivesemileptonicdecays34}

The decay $B^- \to \tau^- \bar \nu_\tau$ has been the first purely
leptonic B decay to be observed \cite{Ikado:2006un}.
In
the absence of new physics, $B  \to l \nu_l$  decays  $ (l= e, \mu,  \tau )$  are simple tree-level decays,
where  the two quarks in the initial state, b and $\bar u$, annihilate to a $W^-$ boson.
Thay are particularly sensitive to physics beyond the SM, since a new particle,
 for example, a
charged Higgs boson in supersymmetry or a
generic two-Higgs doublet model, may lead the decay taking
the place of the $W^-$ boson.
In the SM, the  $B^-  \to \tau^- \bar \nu_\tau$  branching ratio is
\beq {\cal{B}}(B^-  \to \tau^- \bar \nu_\tau) = \frac{G_F^2 m_B m_\tau^2}{8 \pi} \left(1- \frac{m_\tau^2}{m_B^2} \right)^2 f_B^2 |V_{ub}|^2 \tau_B
\eeq
 and its  measurement  provides a direct experimental determination of the
product  $f_B  |V_{ub}|$.
Experimentally, it is challenging to identify the $B^- \to \tau^-  \bar \nu_\tau$   decay because it involves more than one neutrino
in the final state and therefore cannot be kinematically
constrained. At $B$ factories, one can reconstruct
one of the $B$ mesons in the $e^+ e^- \to \Upsilon (4S) \to \bar B B$ chain,
 either
in hadronic decays or in semileptonic decays. One
then compares properties of the remaining particle(s) to those expected for
signal and background.
In contrast with previous data, the new Belle result \cite{Adachi:2012mm}
\beq {\cal{B}}( B^- \to \tau^- \bar\nu_\tau) =( 0.72^{+0.27}_{-0.25}\pm 0.11) \times 10^{-4}
\eeq
where the first errors are statistical and the second ones systematical,
is  in substantial agreement with the SM predictions.
The amount of the agreement varies if we compare with predictions based on specific or averaged $|V_{ub}|$ exclusive and inclusive determinations, indirect $|V_{ub}|$ fits, or estimates where  the dependency
from $|V_{ub}|$ is eliminated, e.g.  by using the unitarity conditions of the
CKM matrix \cite{Lunghi:2010gv}.

\subsection{Inclusive $|V_{ub}|$}
\label{vuninclusivo}

The extraction of $|V_{ub}|$ from inclusive decays requires to address theoretical issues absent in the inclusive $|V_{cb}|$ determination, as outlined in Sect. \ref{subsectionInclusive decays}.
On the experimental side, efforts have been made
 is to enlarge
the  experimental range, so as to reduce,
on the whole,  the weight of the endpoint region.
 Latest results by Belle \cite{Urquijo:2009tp}
 access $\sim  90$\% of the $ \bar B \rightarrow X_u  l \bar \nu_l$ phase space, claiming an overall uncertainty of 7\% on $|V_{ub}|$.
A similar portion of the phase space is covered also by the most recent BaBar analysis \cite{Lees:2011fv}.
From the theoretical side, several available theoretical schemes are available. All of them are  tailored
to analyze data in the threshold region,  but  differ significantly
in their treatment of perturbative corrections and the
parameterization of non-perturbative effects.

The
average values for $|V_{ub}|$ have  been extracted  by HFAG   from the partial branching fractions, adopting
a specific theoretical framework and taking into account
 correlations among the various measurements
and theoretical uncertainties \cite{Amhis:2012bh}.
The latest experimental analysis, Ref. \refcite{Lees:2011fv},  and  the HFAG averages in Ref. \refcite{Amhis:2012bh}
rely on at least four different QCD calculations of the partial
decay rate: BLNP
by Bosch, Lange, Neubert, and Paz \cite{Lange:2005yw, Bosch:2004th, Bosch:2004cb}; DGE, the
dressed gluon exponentiation, by Andersen and Gardi \cite{Andersen:2005mj}; ADFR by Aglietti, Di Lodovico, Ferrara, and Ricciardi
\cite{Aglietti:2004fz, Aglietti:2006yb,  Aglietti:2007ik}; and GGOU by Gambino, Giordano, Ossola
and Uraltsev \cite{Gambino:2007rp}.
 These  QCD theoretical calculations are the ones taking into account  the whole set of experimental results, or most of it, starting from 2002 CLEO data \cite{Bornheim:2002du}.
They can be roughly  divided into approaches based on the estimation of the shape function (BLN, GGOUP) and on resummed perturbative QCD (DGE, ADFR).
Other theoretical schemes have been described  in Refs.  \refcite{Bauer:2001rc},\refcite{Leibovich:1999xf}, \refcite{Lange:2005qn}.

The shape function approach is based on the introduction of a nonperturbative
distribution function (shape function) that at leading order is
universal. The shape function takes care of singular terms in the theoretical
spectrum; it has the role of a momentum distribution function of the $b$-quark
in the $B$ meson. However, the OPE does not predict
the shape function and an ansatz is needed for its functional form. The
subleading shape functions are difficult to constrain and are not process
independent.

Predictions based on resummed perturbative QCD use resummed perturbation
theory in moment space to provide a perturbative calculation of the
on-shell decay spectrum in the entire phase space. They extend the standard
Sudakov resummation framework by adding non-perturbative corrections in
the form of power corrections, whose structure is determined by renormalon
resumming  \cite{Andersen:2005mj} or by an effective QCD coupling \cite{ Aglietti:2004fz, Aglietti:2006yb,  Aglietti:2007ik}. The shape of the
spectrum in the kinematic region, where the final state is jet-like, is largely
determined by a calculation, and less by parametrization. In principle, there
is no preclusion to why an effective coupling inserted in the perturbative resumming
formula cannot adequately describe the non-perturbative Fermi
motion as well as a fitting function (see e.g. Ref. \refcite{Shirkov:1997wi}). In ADFR,  the physical picture implied is that
$B$ fragmentation into the $b$-quark and the spectator quark can be described
as a radiation process off the $b$-quark with a proper coupling. This effective
determination of $|V_{ub}|$ from semileptonic $B$ decays
 is universal in the sense that describes radiative decay processes as
well as $B$ fragmentation processes; once it is fixed, for instance on the basis
of minimal analyticity arguments, there are no free parameters to be fitted
in the model.

Although conceptually quite different, all the above approaches generally
lead to roughly consistent results when the same inputs are used and the
theoretical errors are taken into account.
 Recent HFAG estimates \cite{Amhis:2012bh} are reported in Table \ref{phidectab3} and plotted in  Fig.
\ref{Vub-fig}.
They give values in the range $\sim (3.9-4.6) \times 10^{-3}$; let us observe that
the theoretical uncertainty among determinations can reach 10\%.
%%%%%%%%%%%%%%
\begin{table}[h]
\tbl{Inclusive $|V_{ub}|$ averages \cite{Amhis:2012bh}}
{\begin{tabular*}{0.4\textwidth}{@{\extracolsep{\fill}}cc@{}} \toprule
{\it \bf Theory}  &   $|V_{ub}| \times 10^{3}$ \\
%\hline
%\hline
\colrule
BLNP  & $ 4.40 \pm 0.15^{+0.19}_{-0.21}  $\\
%\hline
DGE   & $4.45 \pm 0.15^{+ 0.15}_{- 0.16}$\\
%\hline
ADFR   & $4.03 \pm 0.13^{+ 0.18}_{- 0.12}$\\
%\hline
GGOU   & $4.39 \pm  0.15^{ + 0.12}_ { -0.20} $\\  \botrule
%\hline
%\hline
\end{tabular*}}
\label{phidectab3}
\end{table}
%%%\end{tabular}
%%%\end{ruledtabular}
%%%\end{table}
%%%\end{tabular}
%%%\end{ruledtabular}
%%%\end{table}
%%%%%%%%%%%%%%%%%%%%%%%%%%%%%%%%%%%%%%

\section{$|V_{tb}|$}
\label{vtb}

The $t$-quark
decays before it can hadronize, since its lifetime $\tau \simeq (1.5$ GeV)$^{-1}$ is  much less than the QCD scale $\simeq (200$ MeV)$^{-1}$; there are no top mesons or baryons.
Information on the value of $|V_{tb}|$ has been traditionally obtained indirectly,
analyzing
loop-dominated observables  sensitive to $|V_{tb}|$, e.g., $B_d$ and $B_s$  mixing or the radiative decay $b \to s \gamma$. The term indirect emphasizes that
 in order to  to extract the desired information one has to consider loop processes and/or make some usage of  SM
properties, such as CKM unitarity.
In the SM,
unitarity constrains $|V_{tb}|$ to be very
close to one  \cite{CKMfitter} \beq  |V_{tb}| = 0.999142^{+0.000043}_{-0.000025} \label{SMvtb} \eeq

An indirect measurement of $|V_{tb}|$ that does not require the assumption of unitarity has been
performed on electroweak data from LEP, SLC, the Tevatron, and neutrino experiments.
The result mostly comes from two-loop contributions to $ \Gamma(Z \to b \bar b)$
and yields \cite{Swain:1997mx}
\beq
|V_{tb}| = 0.77^{+0.18}_{-0.24}
\eeq
At hadron machines,  the top quark is mainly produced in top-antitop pairs via strong
interactions. However, the pure electroweak production of a single top (or anti-top) quark
has a remarkably competitive cross-section.
A test of $|V_{tb}|$
can be made from the measurement of ${\cal{R}} ={ \cal{B}} (t \to W b)/{ \cal{B}}  (t \to W q)$ where ${ \cal{B}}  (t \to W q)$ is
the branching fraction of the top quark to a W boson and a quark ($q$ = $b$, $s$, $d$). This quantity
has been measured at the Tevatron. The latest results, from the \D0
collaboration \cite{Abazov:2011zk}, are  ${\cal{R} }= 0.90 \pm 0.04$ (stat.+syst.),   which agrees within approximately
2.5 standard deviations with the SM prediction of $\cal R$ close to one.
A simultaneous measurement
of ${\cal R} = 0.94 \pm 0.09$ and $\sigma_{\bar t t}$ has been recently performed by CDF;
%assuming three generations
%of quarks and given the unitarity of the CKM matrix, 
they found \cite{Aaltonen:2013doa}
\beq
|V_{tb}| = 0.97 \pm 0.05
\eeq

The single top quark production cross section is directly proportional to the square of $|V_{tb}|$, allowing a direct measurement  of  $|V_{tb}|$ without assuming unitarity of the CKM matrix or three fermion generations.
The  top quark was discovered at Tevatron  in
1995, but,  due to the very low signal-over-background,  it was possible to see  single
top at the Tevatron only in 2009 \cite{Abazov:2009ii,Aaltonen:2009jj}, 14 yars later. Instead,  ATLAS and CMS,
thanks to the much larger cross sections and better signal-over-background
available at the LHC, observed  single top already
in 2011 and measured its cross section the year after.
The current single top cross section measurements,
 (including the
most recent one by CMS at 8 TeV) have uncertainties at
the level of 10\% \cite{Chiarelli:2013psr}, too large to challenge the SM. The only exception is the
 CMS 7  TeV measurement \cite{Chatrchyan:2012ep}, with
uncertainty of $\sim 5$\%, which yields
\beq
|V_{tb}| = 1.02 \pm 0.05 \pm 0.02
\eeq
Due to the large LHC statistics, these measurements are
(mostly) systematics limited.
Dedicated strategies need to  be developed to increase precision and usefully employ this process
in the NP search.
A deviation from the SM prediction in Eq. \ref{SMvtb} could arise from NP contributions that violate unitarity.
Possibly the simplest  way to violate unitarity is enlarging the fermion sector, by including a fourth quark generation or
 vector-like quarks, that appear in many models,  Randall-Sundrum or E6 GUTs amongst others (see, e.g.
  \cite{Botella:2012ju,  Buras:2009ka, Alok:2010zj, Lacker:2012ek, Aguilar-Saavedra:2013qpa}).

\bibliographystyle{ws-mpla}
\bibliography{VxbRef}

\end{document}